\documentstyle[12pt]{article}

\begin{document}

\begin{center}
NON-SINGULAR CONSTANT CURVATURE TWO-DIMENSIONAL BLACK HOLE \\
\vskip 1cm
{\bf Jos\'e P. S. Lemos} \\
\vskip 0.3cm
{\scriptsize Departamento de Astrof\'{\i}sica,
             Observat\' orio Nacional-CNPq,} \\
{\scriptsize  Rua General Jos\'e Cristino 77,
              20921 Rio de Janeiro, Brazil,} \\
{\scriptsize  \&} \\
{\scriptsize  Departamento de F\'{\i}sica,
              Instituto Superior T\'ecnico,} \\
{\scriptsize  Av. Rovisco Pais 1, 1096 Lisboa, Portugal.} \\
\vskip 0.6cm
{\bf Paulo M. S\'a} \\
\vskip 0.3cm
{\scriptsize  Centro de F\'{\i}sica da Mat\'eria Condensada,} \\
{\scriptsize  Av. Prof. Gama Pinto 2,
              1699 Lisboa Codex, Portugal.}
\end{center}

\vskip 2cm

\begin{abstract}
\noindent
  We show that the constant curvature two-dimensional theory of
  Teitel\-boim and Jackiw admits a black hole solution,
  which is free of spacetime singularities.
  The maximally extended spacetime consists of an infinite chain of
  universes connected by timelike wormholes.
\end{abstract}

\newpage
\noindent
Gravitation in two and three dimensions has shown that the
existence of black holes is not limited to the four-dimensional
world.
Thus, it has surprisingly been found that three-dimensional
General Relativity with a constant negative curvature
admits black hole solutions with mass and angular
momentum [1].
On the other hand,
Perry and Teo [2] have shown that the two-dimen\-sional
(2D) string black hole,
which appears as an exact solution of the
Wess-Zumino-Witten theory,
is free of spacetime singularities.
This also surprising result is in contrast with the corresponding
approximated black hole solution found by imposing the
vanishing of the $\beta$ functions in perturbation theory
[3].
In this case,
as well as in General Relativity,
the black hole is followed by a singularity.
The purpose of this communication is to report that the
two-dimensional constant curvature theory of
Teitelboim and Jackiw [4] have analogous properties.
First, it admits a black hole solution,
and secondly, the black hole
is free of spacetime singularities.
This theory is constructed from the action,
\begin{equation}
    S={\frac{1}{2\pi}}\int{ d^2x \sqrt{-g}
       \Phi \left( R + 4\lambda^2\right)},
\label{eq:1}
\end{equation}
where $R$ is the scalar curvature,
$\Phi$ is a scalar field
and $\lambda$ is a constant.
Variation of this action with respect to $g_{ab}$ and $\Phi$
yields  the gravitational and dilaton field equations, respectively,
\begin{eqnarray}
& &   \nabla_a \nabla_b \Phi
     -g_{ab} \nabla_c \nabla^c \Phi
     +2 \lambda^2 g_{ab} \Phi =0,
\label{eq:2}   \\
& &   R + 4\lambda^2=0,
\label{eq:3}
\end{eqnarray}
where $\nabla$ represents the covariant derivative.
In order to find the black hole solution we write the metric
in the unitary gauge
\begin{equation}
  ds^2 = - e^{2\nu} dt^2 + dx^2.
\label{eq:4}
\end{equation}
We now assume that $\nu=\nu(x)$.
Then,
a solution of equations (\ref{eq:2}) and (\ref{eq:3})
is given by
\begin{eqnarray}
 & & \Phi=\Phi_0 \cosh \left( ax \right),
\label{eq:5} \\
 & & ds^2 = -\sinh^2 \left( ax \right) dt^2
            +dx^2,
\label{eq:6}
\end{eqnarray}
where $-\infty<x<+\infty$ and $a=\sqrt{2\lambda^2}$.
Metric (\ref{eq:6}) has a coordinate singularity at $x=0$,
which as we will show corresponds to a horizon.
In order to elucidate the metric,
we transform to the Schwarzschild gauge by
doing
$ar=\cosh \left( ax \right)$,
thus obtaining
\begin{eqnarray}
& & \frac{\Phi}{\Phi_0}=ar,
      \label{eq:8}  \\
& & ds^2= - \left( a^2 r^2 - 1 \right) dt^2
        + \frac{dr^2}{a^2 r^2 - 1}.
      \label{eq:7}
\end{eqnarray}
Note that the line $-\infty<x<+\infty$
corresponds to the segment
$1<r<+\infty$;
each pair of space inverted points $(-x,x)$,
degenerates into just one $r$.
There is a horizon at $ar=1$, i.e.\ $x=0$.
This means that observers at each end of the line,
$x\rightarrow\pm\infty$,
can only communicate if they enter into $x=0$.
In order to show that $x=0$ is a horizon and that solution
(\ref{eq:8})-(\ref{eq:7}) indeed represents a black hole,
we perform the maximal analitical extension.
The connection to Kruskal coordinates is,
$U=-e^{-at} \sqrt{\frac{ar-1}{ar+1}}$,
$V= e^{ at} \sqrt{\frac{ar-1}{ar+1}}$.
Now, spacetime is not singular since the scalar
curvature is a constant.
In figure~1 we draw the Penrose diagram.
Region I contains infinity, $ar=+\infty$
($UV=-1$), represented by a timelike line.
The horizons are at $ar=1$ ($UV=0$).
At $ar=0$ ($UV=1$),
in region II, there is nothing special happening,
no singularity, no horizon, no infinity.
Thus one can continue the solution to negative $r$ values
(using another coordinate system, $\bar{U}=U^{-1}$ and
$\bar{V}=V^{-1}$),
past a new horizon at $ar=-1$
($\bar{U}\bar{V}=0$),
reemerge in region III
(degenerated in the same sense as region I)
which contains the timelike infinity
$ar=-\infty$
($\bar{U}\bar{V}=-1$), and so on.
One therefore has an infinite chain of universes
connected by timelike wormholes.

We note that equations (\ref{eq:2})-(\ref{eq:3})
in the unitary gauge
admit also the anti-de~Sitter solution,
\begin{eqnarray}
 & & \Phi=\Phi_0 \sinh \left( ax \right),
\label{eq:9} \\
 & & ds^2 = -\cosh^2 \left( ax \right) dt^2
            +dx^2,
\label{eq:10}
\end{eqnarray}
where $-\infty<x<+\infty$.
Going to the Schwarzschild gauge,
the metric is
\begin{equation}
  ds^2= - \left( a^2 r^2 + 1 \right) dt^2
        + \frac{dr^2}{a^2 r^2 + 1},
\label{eq:11}
\end{equation}
and $\Phi$ is given as in (\ref{eq:8}).
Since both solutions (\ref{eq:7}) and
(\ref{eq:11}) are of constant curvature and
admit asymptotically ($ar\rightarrow\infty$)
the same isometry group $SO(2,1)$,
their properties can only be distinguished globally.
Indeed, regions I, II and III of figure~1 can be obtained by
cutting pieces of the anti-de~Sitter spacetime
and identifying points properly
in a process similar to that given in ref.~[1].
It is worth to emphasize that the unitary gauge gives
at once the black hole solution,
since the identifications are automatically performed.

An important quantity is the
ADM total mass M.
For the black hole given in equations
(\ref{eq:5})-(\ref{eq:6}) one has
\begin{equation}
    M=\frac{|\lambda|}{2\sqrt2} \Phi_0.
\label{eq:12}
\end{equation}

A more general action is provided by the expression
\begin{equation}
S=\frac{1}{2\pi} \int{ d^2x \sqrt{-g}
   \Phi \left( R - \omega
        \frac{\left( \nabla \Phi \right)^2}{\Phi^2}
        + 2 \lambda^2 \right) },
\label{eq:13}
\end{equation}
with a parameter $\omega$
(usually one puts $\Phi=e^{-2\phi}$ in eq.\ (\ref{eq:13})).
For $\omega=0$ one gets the Teitelboim-Jackiw theory [4],
for $\omega=-\frac12$ one has planar General Relativity [5],
for $\omega=-1$ one has the first order string theory [3],
and for $\omega=\pm\infty$ one gets the 2D analogue of General
Relativity [6].
The black hole solutions of this general theory have been
studied in ref.~[7].

\vskip 1cm

\noindent
{\bf Acknowledgements}

\noindent
JPSL acknowledges grants from JNICT (Portugal) and CNPq (Brazil).
PMS acknowledges a JNICT (Portugal) grant BIC/776/92.
\vskip 1cm

{\bf References}

\begin{enumerate}
\item      M. Ba\~nados, M. Henneaux, C. Teitelboim and J. Zanelli,
           {\it Phys. Rev. D} {\bf 48} (1993) 1506.
\item      M.\ J.\ Perry and  E.\ Teo,
           {\it Phys.\ Rev.\ Lett.} {\bf 70} (1993) 2669.
\item      G. Mandal, A. M. Sengupta and S. R. Wadia,
           {\it Mod. Phys. Lett. A} {\bf 6} (1991) 1685;
           E. Witten,
           {\it Phys. Rev. D} {\bf 44} (1991) 314.
\item      C.\ Teitelboim,
           in: Quantum Theory of Gravity, essays in honour of
           the 60th birthday of B.\ DeWitt, ed.\ S.\ Christensen
           (Adam Hilger-Bristol, 1984) p.\ 327;
           R.\ Jackiw,
           in: Quantum Theory of Gravity, essays in honour of
           the 60th birthday of B.\ DeWitt, ed.\ S.\ Christensen
           (Adam Hilger-Bristol, 1984) p.\ 403.
\item      J.\ P.\ S.\ Lemos,
           ``Two-Dimensional Black Holes and General Relati\-vity'',
           IST Lisbon preprint DF/IST-13.93 (1993).
\item      J.\ P.\ S.\ Lemos and P.\ M.\ S\'a,
           IST Lisbon preprint (1993), in preparation.
           The effective action derived in this limit
           ($\omega\rightarrow\pm\infty$) is equal to the action
           given in
           C. G. Torre,
           {\it Phys. Rev. D} {\bf 40} (1989) 2588;
           R. B. Mann, S. M. Morsink, A. E. Sikkema and T. G. Steele,
           {\it Phys. Rev. D} {\bf 43} (1991) 3948.
\item      J.\ P.\ S.\ Lemos and P.\ M.\ S\'a,
           ``The Black Holes of a General Two-Dimensional Dilaton
           Gravity Theory'',
           IST Lisbon preprint DF/IST-9.93 (1993).
\end{enumerate}

\newpage

{\bf Figure Captions}

\vskip 1cm

Figure 1: Penrose diagram for the maximally extended black hole
          of the Teitelboim-Jackiw theory. There is an infinite
          chain of regions none of them contains a singularity.
\end{document}